\documentclass[pdflatex,sn-mathphys-num]{sn-jnl}
\usepackage{multirow}%
\usepackage{amsmath,amssymb,amsfonts}%
\usepackage{amsthm}%
\usepackage[title]{appendix}%
\usepackage{xcolor}%
\usepackage{textcomp}%
\usepackage{manyfoot}%
\usepackage{booktabs}%
\usepackage{algorithm}%
\usepackage{algorithmicx}%
\usepackage{algpseudocode}%
\usepackage{listings}%

\usepackage{graphicx}
\usepackage{dcolumn}
\usepackage{bm}

\DeclareMathOperator{\arctantwo}{arctan2} 
\DeclareMathOperator{\sgn}{sgn}
\usepackage{textcomp} 
\usepackage{gensymb}
\usepackage[utf8]{inputenc}
\DeclareUnicodeCharacter{00B0}{\degree}
\usepackage{mathrsfs}

\usepackage[percent]{overpic}
\usepackage{pict2e}
\usepackage[dvipsnames]{xcolor}
\definecolor{myGreen}{RGB}{10, 130, 10}

\newcommand{\Grf}{G_\mathrm{rf}}

\usepackage{xcolor}
\usepackage[normalem]{ulem} 

\begin{document}

\title[Phase projection errors in rf-driven optically pumped magnetometers]{Phase projection errors in rf-driven optically pumped magnetometers}

\author*[1]{\fnm{Zoran D.}\sur{Grujić}}
\email{zoran.grujic@ipb.ac.rs}

\author[1]{\fnm{Marija} \sur{Ćurčić}}%

\author[1]{\fnm{Aleksandra} \sur{Kocić}}%

\author[2]{\fnm{Antoine} \sur{Weis}}

\author*[3]{\fnm{Theo} \sur{Scholtes}}
\email{theo.scholtes@leibniz-ipht.de}

\affil[1]{\orgdiv{Institute of Physics Belgrade}, \orgname{University of Belgrade}, \orgaddress{\street{Pregrevica 118}, \postcode{11080} \city{Belgrade}, \country{Serbia}}}

\affil[2]{\orgdiv{Physics Department}, \orgname{University of Fribourg},\orgaddress{ \street{Chemin du Musée 3}, \postcode{1700} \city{Fribourg}, \country{Switzerland}}}

\affil[3]{\orgname{Leibniz Institute of Photonic Technology}, \orgaddress{\street{Albert-Einstein-Strasse 9}, \postcode{07745} \city{Jena}, \country{Germany}}}


\abstract{
We investigate the phase relationship between the oscillating (rf) excitation field and the detected (light) power modulation in scalar rf-driven optically pumped magnetometers (OPMs), in particular in the $M_x$ configuration. 
While the static dependence of the demodulation phase on the direction of the external static magnetic field vector can be largely mitigated by aligning the oscillating rf field along the light propagation direction, we demonstrate that a dynamic (transient) phase response arises under magnetic field tilts.
We analytically solve the corresponding modified Bloch equation and confirm agreement with experimental observations obtained using an $M_x$ magnetometer incorporating a paraffin-coated Cs vapor cell.
The results reveal fundamental limitations of $M_x$ magnetometers regarding response time and accuracy, in particular when employed with active electronic feedback, such as a phase-locked loop.
Therefore, this work is highly relevant to important magnetometry applications where the direction of the quasi-static magnetic field of interest is unknown \textit{a priori} or varies over time, or in measurements requiring a large detection bandwidth.
Such conditions are encountered in applications such as geomagnetic surveying, particularly with mobile platforms.
}


\maketitle

\section{\label{sec:introduction}Introduction}
Optically pumped magnetometers (OPM) represent the most sensitive non-cryogenic magnetic field sensors currently available \cite{Budker2013}.
Their exquisite performance is crucial in applications requiring the detection or monitoring of extremely small magnetic field changes, such as magnetocardiography \cite{Bison2009, Morales2017,Jensen2018,Yang2021}, magnetoencephalography \cite{Boto2018,Brookes2022}, fundamental physics experiments \cite{GNOME2021,nEDM2020}, geomagnetic prospection \cite{Oelsner2022,Lu2023}, detection and imaging of magnetic nanoparticles \cite{Jaufenthaler2020,Lebedev2022,Sasayama2024}, exploration of magnetic fields in space \cite{Pollinger2020,Schirninger2021,Deans2023}, and many others, as recently reviewed in \cite{Fu2020}. 

OPMs are based on a variety of magneto-optical effects in (alkali-metal) vapors \cite{Budker2002} and are implemented in a wide range of implementations tailored to specific applications.

The $M_x$ magnetometer, with a long history of successful operation, is a proven choice for both sensitivity and robustness \cite{BellBloom1957,Bloom1962,Aleksandrov1995,Groeger2006,Schwindt2007,Bison2009,Scholtes2011}.
In this type of OPM, the modulus $B_0$ of the static magnetic field vector $\vec{B}_0$ is inferred from the driven Larmor precession frequency, $\omega_\mathrm{L}= \gamma_{F} \, B_0$, of an ensemble of polarized gaseous atomic spins, typically enclosed within a vapor cell (for ${}^{133}\mathrm{Cs}$ used here, $\gamma_F/2\pi \approx 3.5 \,\mathrm{Hz/nT}$).
The Larmor precession is 'driven' by an additionally applied weak magnetic field oscillating at $\omega_\mathrm{rf}$ near the Larmor frequency.
In the standard $M_x$ magnetometer a single circularly polarized, light beam propagating along $\vec{k}$ and tuned resonantly to an atomic transition (typically the $D_1$ and $D_2$ lines in alkali atoms) is used for both, the creation of atomic spin polarization by a process called optical pumping \cite{Happer1972}, and the read-out of the spin precession signal. 
As the light beam passes through the vapor cell, it acquires a modulation at the rf field frequency, which can be detected by a photodiode placed behind the cell.
The amplitude and phase of this light modulation exhibit resonant behavior when the rf field frequency and Larmor frequency coincide.
The steady-state phase signal, typically recorded using a lock-in amplifier (see Sec.~\ref{sec:experimentalSetup}), is described by \cite{Colombo2017}
\begin{equation}\label{eq:phase}
\varphi(\delta\omega)=\varphi^0 - \arctan{\frac{\delta\omega}{\gamma}}\,,
\end{equation}
where $\delta\omega=\omega_\mathrm{rf}-\omega_\mathrm{L}$ is the rf detuning and $\gamma$ the spin relaxation rate. 
When $\omega_\mathrm{rf}$ is fixed, a change in the modulus of the static magnetic field $|\vec{B}_0|$ will lead to a change in $\delta\omega$ and thus in $\varphi$.
For small changes in the magnetic field modulus (i.e. when $\delta\omega$ is small compared to $\gamma$), the sensor response is linear and can be used in a 'free-running' mode to monitor field changes.
However, to track larger changes in the Larmor frequency, active feedback is commonly employed. 
The phase signal is particularly well-suited for this purpose due to its reduced sensitivity to technical noise, compared to in-phase, quadrature, and magnitude lock-in signals, and its effective suppression of rf broadening \cite{Bison2004}.
In this approach, the feedback typically adjusts $\omega_\mathrm{rf}$ to maintain $\varphi$ at a constant setpoint, often choosing $\varphi^0$ at the point of maximum signal slope.
This mode of operation, often referred to as a 'phase locked loop' (PLL) can increase both the dynamic range and measurement bandwidth \cite{MagResMag}.
However, for satisfactory operation of either method, the offset phase $\varphi^0$ must be extremely carefully controlled. 
Disregarding electronic phase shifts \footnote{Electronic (possibly frequency-dependent) phase shifts can arise from rf coil and cable inductance, limited bandwidth of the photo current transimpedance amplifier, etc.}, $\varphi^0$ depends on the orientation of $\vec{B}_\mathrm{rf}$ and $\vec{B}_0$ relative to the light propagation direction $\vec{k}$.
In the classical $M_x$ magnetometer implementation \cite{BellBloom1957,Bloom1962,Aleksandrov1995,Groeger2006,Bison2009}, where $\vec{B}_\mathrm{rf}\perp\vec{k}$, the offset phase is given by
\begin{equation}
    \varphi^0_\perp = \arctan({\cos{\alpha}\cot{\phi}})\,,
\end{equation}
where $\alpha$ is the angle between $\vec{B}_0$ and $\vec{k}$ and $\phi$ is the angle between $\vec{B}_\mathrm{rf}$ and the projection of $\vec{B}_0$ into the plane perpendicular to $\vec{k}$ \cite{Colombo2017}.
Consequently, a change in direction of $\vec{B}_0$ does in general lead to a change in $\varphi^0_\perp$, and thus in $\varphi$.
Hence, when the phase signal is used within an active feedback stabilizing $\varphi$, a change in direction of $\vec{B}_0$ can mimic a change in its magnitude, leading to an erroneous feedback correction to $\omega_\mathrm{rf}$.
We term this effect the 'static phase projection error' (SPPE).
Thus, in this configuration, the feedback loop can only operate effectively only if the direction of $\vec{B}_0$ remains constant over time relative to $\vec{k}$ and $\vec{B_\mathrm{rf}}$.

When the sensor geometry is adjusted to have $\vec{B}_\mathrm{rf}\parallel\vec{k}$, it has been shown in \cite{Colombo2017}, that the offset phase is constant for all angles $\alpha$ and $\phi$, except for a $\pi$ phase shift when transitioning between the two half-spaces with either $\vec{k}$ parallel or anti-parallel to the projection of $\vec{B}_0$:
\begin{equation}\label{eq:phizero}
    \varphi^0_\parallel = -\frac{\pi}{2}\mathrm{sign}({\cos{\alpha}})\,.
\end{equation}
Because the offset phase $\varphi^0_\parallel$ remains constant within each half-space, no phase change should occur as the direction of $\vec{B}_0$ varies within it.
Consequently, this configuration has been termed the \textit{true-scalar magnetometer} (TSM) \cite{MagResMag}.

While the TSM geometry elegantly eliminates the SPPE under steady-state conditions, its immunity has never been tested rigorously for time-dependent rotations of $\vec{B}_0$. 

Most of the practical applications, particularly airborne, maritime, mobile geomagnetic mapping and wearable biomedical systems, experience continuous reorientation of the ambient magnetic field.
A full understanding of the transient effects (beyond the steady-state treatment) is therefore essential for a reliable operation of OPMs under realistic conditions.

In this paper, we demonstrate that Eq.~\ref{eq:phizero} holds true only within the steady-state approximation used in its derivation.
In the general case, when $\vec{B}_0$ changes direction, a transient phase change is observed, which, after a time interval scaling with $\gamma$, settles back to the expected steady-state value.
We term this effect 'dynamic phase projection error' (DPPE) and its understanding is the primary motivation of this work.
The nature of DPPE depends strongly on the direction of the $\vec{B}_0$ rotation, with two distinct cases: rotation within the plane spanned by $\vec{B}_0$ and $\vec{k}$ and rotation of $\vec{B}_0$ out of this plane.
As we show below, in the latter case, a residual phase shift remains even when averaged over one or more rf cycles, as is typically done in lock-in detection.

To the best of our knowledge, this work provides the first explicit analytical and experimental characterization of such transient phase errors in a TSM geometry.
This fills an important gap in the modeling of rf-driven OPMs, where transient orientation dynamics are generally neglected.
The results have direct implications for a PLL-based operation, heading-error suppression strategies and the design of next-generation of the scalar OPMs that are deployed on moving platforms or in dynamically varying fields.

The remainder of this paper is organized as follows:
In Sec.~\ref{sec:theory}, we solve the equation of motion for the spin orientation vector, obtaining analytical expressions in the resonant case ($\delta\omega=0$) that reproduce the DPPE phenomenon. 
In Sec.~\ref{sec:experiment}, we detail the experiments conducted to demonstrate the effect and present the experimental results in agreement with model predictions. 
Finally, we discuss implications of the effect for $M_x$ magnetometers in real-world applications in Sec.~\ref{sec:discussion} and conclude the paper in Sec.~\ref{sec:conclusion}.
\section{\label{sec:theory}Theory}
We extend previous work by Weis et~al.~\cite{MagResMag} by solving the modified Bloch equation in the case $\vec{B}_\mathrm{rf}\parallel\vec{k}$ (TSM), including the transient evolution of the spin orientation vector and deriving equations describing the signals detected by a lock-in amplifier in the experiment.
We provide the analytical solution for the case of a resonant rf magnetic field ($\delta\omega=0$).

We establish the modified Bloch equation in the laboratory frame, solve it in the rotating frame using the rotating wave approximation (RWA), transform the solution back into the laboratory frame, and obtain the lock-in signals.
Finally, to analyze the DPPE, we consider rotations around the $x$- and $y$-axes, respectively.

\subsection{The Bloch equation}
\label{sec:Bloch}
The equation of motion we consider here is the Bloch equation \cite{Bloch1946} extended to include optical pumping \cite{MagResMag}.
It describes the evolution of the atomic spin orientation vector $\vec{S}(t)$ under the influence of the magnetic field $\vec{\Omega}(t) = \gamma_F \vec{B}(t)$ and resonant circularly polarized light propagating along $\vec{k}$, resulting in optical pumping at a rate $\gamma_\mathrm{p}$: 
%
%
\begin{eqnarray}
	\label{eq:Bloch}
	\dot{\vec{S}}(t) &=& \vec{S}(t) \times \vec{\Omega}(t) - \gamma \vec{S}(t) + \gamma_\mathrm{p}\left(\vec{k}- \vec{S}(t)\right) .
\end{eqnarray}
The first term $ \vec{S}(t) \times \vec{\Omega}(t)$ describes the torque of the total magnetic field exerted on $\vec{S}$ where 
\begin{eqnarray}
	\label{eq:omegaL}
	\vec{\Omega}(t) &=& \vec{\Omega}_\mathrm{L} + \vec{\Omega}_\mathrm{rf}(t)\\
	&=& \gamma_F \left(\vec{B}_0+\vec{B}_\mathrm{rf}(t)\right).
\end{eqnarray}
The constant $\gamma_F$ is the gyromagnetic ratio, $\vec{\Omega}_\mathrm{L} = \gamma_F \vec{B}_0$ is the angular Larmor frequency, with $\vec{B}_0$ the static magnetic field and $\vec{B}_\mathrm{rf}(t)$ the oscillating magnetic field:
\begin{eqnarray}
	2\, \vec{\Omega}_\mathrm{rf}(t) &=&  2\, \vec{\Omega}_\mathrm{rf} \sin(\omega_\mathrm{rf}t)\\
	&=&  \gamma_F \vec{B}_\mathrm{rf}(t)\\
	&=&  \gamma_F \vec{B}_\mathrm{rf} \sin(\omega_\mathrm{rf}t)\,.
\end{eqnarray}

The optical pumping rate $\gamma_\mathrm{p}$ gives rise to the creation of spin orientation along the laser beam propagation direction $\vec{k}$. 
Please note that $|\vec{k}| = 1$ and $|\vec{S}(t)| \leq 1$, so that the pumping process tends to maximize the atomic spin orientation parallel to $\vec{k}$, counteracted by spin relaxation.
For simplicity, we assume isotropic relaxation of $\vec{S}$ that can be expressed as scalar relaxation rate $\gamma$, representing spin relaxation processes like collisions of atoms with the cell wall or other atoms/molecules within the cell \cite{Scholtes2014}.

\subsection{TSM geometry}
\begin{figure}
	\centering
  \includegraphics[width=0.5\columnwidth]
 {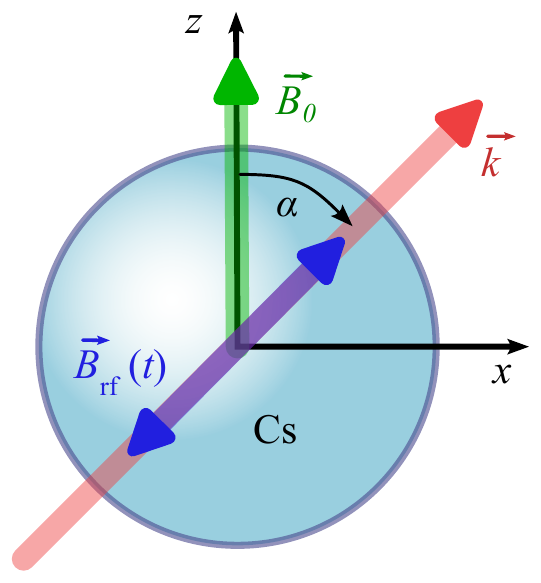}
	\caption{\label{fig:tsm-geometry}
		Geometry of the true scalar magnetometer (TSM).
		The static magnetic field $\vec{B}_0$ is oriented along the $z$-axis, $\vec{k}$ lies in the $zx$-plane, at an angle $\alpha$ to $\vec{B_0}$. The rf magnetic field $\vec{B}_\mathrm{rf}(t)$ oscillates parallel to $\vec{k}$.
	}
	
\end{figure}
The geometry of the TSM is fully defined by two vectors: $\vec{\Omega}_\mathrm{L} = (0, 0, \omega_\mathrm{L})^\top$, where $\omega_\mathrm{L}=\gamma_F |\vec{B}_o |$, and $\vec{k} || \vec{\Omega}_\mathrm{rf}$, where $\vec{\Omega}_\mathrm{rf} = \Omega_\mathrm{rf} (\sin{\alpha}, 0, \cos{\alpha})^\top$. 
For convenience, we choose the $z$-axis to point along the magnetic field, with $\vec{k}$ lying in the $zx$-plane, tilted by an angle $\alpha$ with respect to the $z$-axis, see Fig.~\ref{fig:tsm-geometry}. 
The optimal performance of the $M_x$ magnetometer in terms of sensitivity is achieved at $\alpha \approx 45 \degree$. 
Note that this definition is valid for any orientation of the magnetic field; if the field is not parallel to the $z$-axis, the coordinate system can be rotated to align the field with the $z$-axis and $\vec{k}||\vec{B}_\mathrm{rf}$ in $zx$-plane, resulting in an equivalent TSM magnetometer, with a potentially different angle $\alpha$.
\subsection{Rotating wave approximation (RWA)}
\label{sec:rwa}
To solve the Bloch equation, we transform into the rotating frame (rotating at $\omega_\mathrm{rf}$ in the clockwise (cw) direction), apply the rotating wave approximation (RWA), and transform the solution back into the laboratory frame.
We choose the $z$-axis as our rotation axis, as the spin orientation precesses around it.  
For simplicity, we set $\omega_\mathrm{L} = \omega_\mathrm{rf}$.
In this case, the static magnetic field vanishes in the rotating frame: $\vec{\Omega}_\mathrm{L}^{\mathrm{R}} = (0, 0, 0)^{\top}$.
We use a superscript '$\mathrm{R}$' to denote quantities in the rotating frame, and label the axes in the rotating frame as $\hat{x}$, $\hat{y}$ and $\hat{z}$.
We can split $\vec{\Omega}_\mathrm{rf}$ into two components: one in the $z$-direction (which, with a time average of zero, can be disregarded), and the second in the $x$-direction which can be split into two counter-rotating components:
\begin{eqnarray}
	\mathrm{cw} &\quad \rightarrow \quad&   \Omega_\mathrm{rf}  (\sin({\omega_\mathrm{rf} t}), +\cos({\omega_\mathrm{rf} t}), 0)^{\top} \sin{\alpha}\,, \\
	\mathrm{ccw} &\quad \rightarrow \quad&  \Omega_\mathrm{rf} (\sin({\omega_\mathrm{rf} t}), -\cos({\omega_\mathrm{rf} t}), 0)^{\top} \sin{\alpha}\,.
\end{eqnarray}
Now, the essence of the RWA is to disregard components rotating in the opposite direction of the spin, as they have lower impact than the co-rotating components.
The resulting contributions are at frequency of $2 \omega_\mathrm{rf}$, which is far off-resonance and can therefore be neglected.
For a more detailed discussion, we refer the reader to \cite{MagResMag}.
As $\vec{\Omega}_\mathrm{L}^\mathrm{R}$ vanishes, the remaining cw component of the total magnetic field in the rotating frame is equal to the projected rf magnetic field: 
\begin{eqnarray}
	\vec{\Omega}^\mathrm{R} = \vec{\Omega}_\mathrm{rf}^\mathrm{R} = \Omega_\mathrm{rf}
	\begin{pmatrix} 
		0\\
		\sin{\alpha}\\
		0
	\end{pmatrix}.
\end{eqnarray}
Similarly, $\vec{k}$ is decomposed into a fixed component parallel to $z$, and a component along the $x$-direction.
In the rotating frame, the $x$-component of $\vec{k}^\mathrm{R}$ is not static; it rotates in the $\hat{x}\hat{y}$-plane and generates spin orientation in all directions perpendicular to $\hat{z}$.
This averages to zero, and we can write:
\begin{equation}
\label{eq:k-rot}
	\vec{k}^\mathrm{R} =
	\begin{pmatrix}
		0\\0\\ \cos{\alpha}
	\end{pmatrix}
	.
\end{equation}
Finally, the equation of motion in the rotating frame is:
%
\begin{eqnarray}
	\label{eq:BlochR}
	\dot{\vec{S}}^\mathrm{R}(t) &=& \vec{S}^\mathrm{R}(t) \times \vec{\Omega}^\mathrm{R} - \gamma \vec{S}^\mathrm{R}(t) +\! \gamma_\mathrm{p}\!\!\left(\vec{k}^\mathrm{R}- \vec{S}^\mathrm{R}(t)\right)\!,	\\
	\label{eq:BlochR2}
	\dot{\vec{S}}^\mathrm{R}(t) &=& \vec{S}^\mathrm{R}(t) \times \Omega_\mathrm{rf} 
	\begin{pmatrix} 0\\ \sin{\alpha}\\ 0 \end{pmatrix} 
	- \gamma \vec{S}^\mathrm{R}(t) \nonumber 
	+\;\gamma_\mathrm{p}\left( 	\begin{pmatrix} 0\\ 0\\ \cos{\alpha}\end{pmatrix}
	- \vec{S}^\mathrm{R}(t)\right).	
\end{eqnarray} 
%
As this equation is free from time-dependent coefficients, we can solve it analytically.
\subsection{Analytic solution}
\label{sec:solution}
Equation \ref{eq:BlochR2} was solved using \textit{Wolfram Mathematica} \cite{Mathematica} assuming $0<\alpha<\pi/2$, and setting the initial condition to be
\begin{eqnarray}
	\vec{S}^\mathrm{R}(0) &=& \vec{S}(0) = \begin{pmatrix} S_x^0\\ S_y^0\\ S_z^0\end{pmatrix}.
\end{eqnarray}
We split the solution into stationary $\vec{S}^\mathrm{RS}$ and transient $\vec{S}^\mathrm{RT}(t)$ parts: $\vec{S}^\mathrm{R}(t) = \vec{S}^\mathrm{RS} + \vec{S}^\mathrm{RT}(t)\,,$
with
\begin{eqnarray} \label{eq:stac-state}
	\vec{S}^\mathrm{RS} &=& \gamma_\mathrm{p} \, \mathscr{L}   \begin{pmatrix} -\Omega_\mathrm{rf}\sin({2\alpha})\\ 0\\ 2\gamma_\mathrm{tot}  \cos({\alpha}) \end{pmatrix} \,\text{and}\\
	\vec{S}^\mathrm{RT}(t) &=& e^{-\gamma_\mathrm{tot}  t} \; \begin{pmatrix} S_x^\mathrm{RT}(t)\\ S_y^\mathrm{RT}(t)\\ S_z^\mathrm{RT}(t) \end{pmatrix}\,, \text{with}
\end{eqnarray}
%
\begin{eqnarray}
	S_x^\mathrm{RT}(t) &=& 
	\left[ S_x^0  + 2 \mathscr{L} \gamma_\mathrm{p} \Grf\cos({\alpha})  \right] \cos(\Grf \, t)    \nonumber \\
	&+&  \left[ - S_z^0  +  2\mathscr{L} \gamma_\mathrm{p} \gamma_\mathrm{tot}  \cos({\alpha})    \right]
    \sin(\Grf \, t)
	\,,\\
	S_y^\mathrm{RT}(t) &=& S_y^0\,, \\
	S_z^\mathrm{RT}(t) &=& 
 \left[ S_z^0  -2\mathscr{L}\gamma_\mathrm{p} \gamma_\mathrm{tot}  \cos({\alpha})   \right] 
  \cos(\Grf \, t)  \nonumber \\ 
	&+&  \left[ S_x^0   + 2 \mathscr{L} \gamma_\mathrm{p} \Grf\cos({\alpha})   \right]
	\sin(\Grf \, t) 
	 \,,\
\end{eqnarray}
where we substituted 
\begin{eqnarray}
\gamma_\mathrm{tot} &=& \gamma + \gamma_\mathrm{p}\,,\\
G_\mathrm{rf} &=& \Omega_\mathrm{rf}\sin({\alpha})\,,\\
\mathscr{L} &=& \frac{1}{2} \frac{1}{\gamma_\mathrm{tot} ^2 + G_\mathrm{rf}^2}\,.
\end{eqnarray}
Please note here, that the stationary solution $\vec{S}^\mathrm{RS}$ is not a function of time: 
If we use  $\vec{S}^\mathrm{RS}$ as initial condition, the system will not have to evolve towards the steady state as it is already there.

The spin orientation in the laboratory frame $\vec{S}^\mathrm{L}(t)$ is obtained after multiplying the solution in the rotating frame by the inverse of the rotation matrix we used to switch into the rotating frame:
\begin{eqnarray}
	\vec{S}^\mathrm{L}(t) &=& R_z^{-1}(\omega_\mathrm{rf} t + \phi ) \,\vec{S}^\mathrm{R}(t)\,, \\
	R_z^{-1}(\omega_\mathrm{rf}t + \phi) &=& 
	\begin{pmatrix}
		\cos({\omega_\mathrm{rf}t} + \phi) & \sin({\omega_\mathrm{rf}t} +\phi) & 0\\
		-\sin({\omega_\mathrm{rf}t} + \phi) & 	\cos({\omega_\mathrm{rf}t} + \phi) & 0 \\
		0 & 0 & 1
	\end{pmatrix}\,,
\end{eqnarray}
where the superscript '$\mathrm{L}$' denotes the laboratory frame.

Note the additional phase $\phi$ introduced to account for the ambiguity between the projection of $\vec{k}^\mathrm{R}$ onto the $xy$-plane and $\vec{\Omega}^\mathrm{R}$.
This is a consequence of the approximation introduced in Eq.~\ref{eq:k-rot}. 
Also, the rf field will acquire the same phase in the laboratory frame, 
\begin{eqnarray}
	2\, \vec{\Omega}_\mathrm{rf}^\mathrm{L}(t) &=&  \gamma_F \vec{B}_\mathrm{rf} \sin(\omega_\mathrm{rf}t+\phi)\,.
\end{eqnarray}
\subsection{Lock-in signals}
\label{sec:lockinsignals}
In the first-order approximation of the Lambert-Beer law, the detected signal in the laboratory frame $s(t)$ is proportional to the scalar product of $\vec{k}$ and $\vec{S}^\mathrm{L}(t)$ \cite{AMFMPM},
\begin{eqnarray}
	s(t) &\propto& \vec{k} \cdot \vec{S}^\mathrm{L}(t)\,.  
\end{eqnarray}
We decompose the signal $s(t)$ into in-phase ($X$) and in-quadrature ($Y$) components by grouping terms associated with $\sin({\omega_\mathrm{rf}t+\phi})$ and $\cos({\omega_\mathrm{rf}t+\phi})$,
\begin{align}
s(t) \propto & (X^\mathrm{S}+X^\mathrm{T}) \sin({\omega_\mathrm{rf}t+\phi})  \nonumber \\
+&(Y^\mathrm{S} + Y^\mathrm{T}) \cos({\omega_\mathrm{rf}t+\phi})\,,
\end{align}
and obtain
\begin{align}
	X^\mathrm{S} &= 0\,,  \\
	Y^\mathrm{S} &= - \; \gamma_\mathrm{p} \;\mathscr{L} \; \Grf\sin({2\alpha})\,, \\
	X^\mathrm{T} &=  S_y^0 \; \sin({\alpha}) \; e^{-\gamma_\mathrm{tot}  t}\,,\\
	Y^\mathrm{T} &= \sin({\alpha}) 
         \left[ 
         \cos(\Grf t) \left( S_x^0 + 2 \;\mathscr{L} \gamma_\mathrm{p} \Grf \cos\alpha \right) \right .\nonumber \\
         & \left . +\sin(\Grf t) \left( 2 \;\mathscr{L} \gamma_\mathrm{p} \gamma_\mathrm{tot} \cos\alpha - S_z^0 \right)
         \right] e^{-\gamma_\mathrm{tot}  t} \; \,,
\end{align}
where the superscripts '$\mathrm{S}$' and '$\mathrm{T}$' label static and transient parts, respectively.
We disregard the DC offset of the signal as it is irrelevant for further discussion. 

The signal's phase $\varphi$ can be expressed as $\arctan$ of the ratio between quadrature ($Y$) and in-phase ($X$) signals. 
Here, a special attention needs to be taken regarding the sign of both $X$ and $Y$, defining the quadrant of point $(X, Y)$.
In the general case the function atan2 or arctan2 (as defined in most programming languages like Fortran, Python, etc.) can be used to handle the quadrants by always returning the correct value by the definition
\begin{eqnarray}
    \label{eq:atan2}
	\arctantwo(y,x) &=& \left\{
	\begin{array}{ll}
		\arctan\frac{y}{x}						&  x>0 \\
		\frac{\pi}{2} - \arctan\frac{x}{y} 		&  y>0 \\
		-\frac{\pi}{2} - \arctan\frac{x}{y} 	&  y<0 \,\, (\text{our case})\\
		\arctan\frac{y}{x}	\pm \pi				&  x<0 \\
		\mathrm{undefined}						&  x=0 \quad \mathrm{and}\quad y=0
	\end{array}
	\right .
\end{eqnarray}
As $Y<0$ in our case, the phase signal $\varphi(t)$, measured by a lock-in amplifier, can be written as:
\begin{eqnarray}
	\varphi(t) &=& - \frac{\pi}{2} - \arctan \left( \frac{X^\mathrm{S}+X^\mathrm{T}}{Y^\mathrm{S}+Y^\mathrm{T}}\right),\\
	\label{eq:phi}
	&=& - \frac{\pi}{2} - \arctan \left(\frac{  S_y^0 \; e^{-\gamma_\mathrm{tot}  t} \; }
	{ S_x^\mathrm{RT}(t)  e^{-\gamma_\mathrm{tot}  t}
		- 2 \gamma_\mathrm{p} \mathscr{L} \Grf\cos{\alpha} } \right)\,.
\end{eqnarray}
The phase $\varphi(t)$ is equal to $-\pi/2$ in the steady state:
\begin{eqnarray}
	\vec{S}^\mathrm{L}(t \rightarrow \infty) &=& R_z^{-1}(\omega_\mathrm{rf}t + \phi) \vec{S}^\mathrm{RS}\,.
	\label{eq:sl-steady-state}
\end{eqnarray}
\subsection{The Dynamic Phase Projection Error}
If we turn on our TSM at $t=-\infty$ in a constant magnetic field, $\vec{B}_0\neq \vec{B}_0(t)$, it will be in a steady state at $t=0$:
\begin{eqnarray}
	\vec{S}(0) &=& R_z(\phi) \vec{S}^{RS}\,,
\end{eqnarray}
with Eqs.~\ref{eq:stac-state}~and~\ref{eq:sl-steady-state}.
If the modulus of the magnetic field $|\vec{B}_0(t)|$ changes, the Larmor frequency changes instantaneously, and $\varphi$ changes correspondingly, reflecting the difference between the new Larmor frequency $\omega_L$ and the set rf frequency $\omega_\mathrm{rf}$ (Eq.~\ref{eq:phase} \cite{MagResMag}).

However, now we analyze the case when the modulus remains constant, $|\vec{B}_0(t)|=\mathrm{const}$, while the direction of $\vec{B}_0(t)$ changes.
In such a case, $\varphi$ of a TSM should not change; however, as we show, may undergo an unexpected transient evolution, before settling back to its original, expected steady-state value of $\varphi=-\pi/2$.

If a rotation around the $x$- or the $y$-axis by an angle $\beta$, represented by rotation matrix $R_{x,y}(\beta)$, is applied to $\vec{B}_0$ at $t=0$, the geometry of the TSM will change, altering its relation to the $\vec{S}^0$ and potentially making it no longer in the steady state of the modified TSM.

\begin{figure}
	\centering
	\includegraphics[width=0.35\columnwidth]{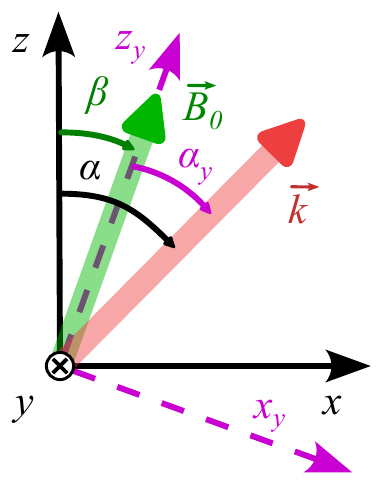}
	\caption{\label{fig:fastv_geometry} Transformation of the coordinate system and magnetic field components for a rotation of $\vec{B}_0$ around the $y$-axis.}
	\label{fig:y-axis-rotation}
\end{figure}

Let us consider a rotation around $y$-axis (Fig.~\ref{fig:y-axis-rotation}). 
The atomic spin will now precess in a plane orthogonal to the new magnetic field vector.
This defines our new $z_y$-axis, with the spin precessing in the $x_yy$-plane.
In this new coordinate system, defined by $x_z,y,z_y$, the TSM operates without further modifications, except for a slightly changed $\alpha$ and modified initial conditions.
Consequently, we have
\begin{eqnarray}
	\alpha_y  &=&  \alpha-\beta\,,\\
	\vec{S}_{y}(0) &=&  R_{y}^{-1}(\beta)\,\vec{S}(0)\,.
\end{eqnarray}

\begin{figure}
	\centering
	\includegraphics[width=0.95\columnwidth]{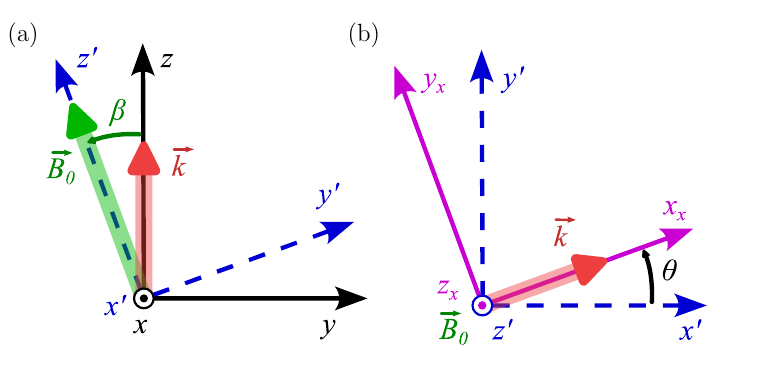}
	\caption{\label{fig:slow_geometry} Transformation of the TSM geometry under a rotation of the static magnetic field $\vec{B}_0$ around the $x$-axis. (a) Initial rotated frame where $\vec{k}$ is misaligned with the TSM plane. (b) Additional rotation restoring $\vec{k}$ to the effective TSM geometry.}
\end{figure}

The $x$-rotation case is more complex, and the transformation will be performed in two steps (Fig.~\ref{fig:slow_geometry}).
First, by $x$-rotation we obtain a $xy'z'$-coordinate system, but as vector $\vec{k}$ is not in the $xz'$-plane, the model we developed for TSM cannot be applied, cf.~Fig.~\ref{fig:slow_geometry}~(a).
Next, we perform a rotation around the $z'$-axis by an angle $\theta$ to bring $\vec{k}$ into the $x_xz_x$-plane, cf.~Fig.~\ref{fig:slow_geometry}~(b).
The detailed description of the transformation is:
\begin{eqnarray}
	\alpha_x  &=&  \arccos{(\cos\alpha\cos\beta)}\approx \alpha\,,\\
	\theta &=& \sgn(\beta) \arccos{\frac{\sin\alpha}{\sqrt{\sin^2\alpha+\cos^2\alpha\sin^2\beta}}}\,, \\
	\theta &\approx& \beta \cot\alpha\,, \label{eq:theta}\\
	\vec{S}_{x}(0) &=&  R_{z'}^{-1}(\theta) R_{x}^{-1}(\beta)\,\vec{S}(0)\,,
\end{eqnarray}
including the approximation for $\theta$ for a small rotation angle $\beta$.
After some calculations we obtain
%
	\begin{eqnarray}
		\label{eq:SLxprime}
		\vec{S}_{x}(0) &=& 2\gamma_\mathrm{p}\, \mathscr{L}\cos\alpha
		\begin{pmatrix}
			\gamma_\mathrm{tot}\sin\phi\sin\beta
			-\Grf\left(\cos\beta\sin^2\phi + \cos^2\phi\right)\\
			\cos\phi \left[ \Grf( cos\beta -1 ) -\gamma_\mathrm{tot} \sin\beta      \right]      \\
			\gamma_\mathrm{tot} \cos\beta + \Grf\sin\phi\sin\beta
		\end{pmatrix},
		\\
		\label{eq:SLyprime}
		\vec{S}_{y}(0) &=& 2\gamma_\mathrm{p}\, \mathscr{L} \cos\alpha
		\begin{pmatrix}
			\gamma_\mathrm{tot} \cos\phi\sin\beta -\Grf(\sin^2\phi+\cos\beta\cos^2\phi)\\
			\sin\phi \left[ \Grf(1-\cos\beta)\cos\phi + \gamma_\mathrm{tot} \sin\beta\right]\\
			\gamma_\mathrm{tot} \cos\beta + \Grf \cos\phi \sin\beta
		\end{pmatrix}.
	\end{eqnarray}
%

%
By taking Eq.~\ref{eq:SLxprime} and \ref{eq:SLyprime} as initial conditions for the phase calculation in Eq.~\ref{eq:phi}, we obtain cumbersome expressions allowing very limited insight and not being suitable for presentation. 
Fortunately, the first order approximation for small angle $\beta$ leads to a compact and insightful form:
\begin{eqnarray}
	\varphi_x(t) &=&   -\frac{\pi}{2} +a_x e^{-\gamma_\mathrm{tot}t}\,,\label{eq:phase_transient}
	\\
	\varphi_y(t) &=& -\frac{\pi}{2} + a_y e^{-\gamma_\mathrm{tot}t} \,,\label{eq:no_phase_transient}\\
	a_x &=& \theta + \beta\frac{\gamma_\mathrm{tot}}{G_\mathrm{rf}}\cos\phi\,, \label{eq:amplitude-x}\\
	a_y &=& - \beta\frac{\gamma_\mathrm{tot}}{G_\mathrm{rf}}\sin\phi \,.  \label{eq:amplitude-y}
\end{eqnarray}
This shows that the TSM is not truly scalar on timescales defined by $\gamma_\mathrm{tot}$.
There is an evolution from one steady state to a new steady state in a slightly changed TSM geometry.
Averaging the phase $\varphi$ over $\phi \in [0,2\pi]$ reveals that in the y-rotation case, there is no shift, while in other case of x-rotation, the shift is $\theta$:
\begin{eqnarray}
	\overline{\varphi\vphantom{A}}_x &=&   -\frac{\pi}{2} +\theta\,,\label{eq:phase-x-transient_average}
	\\
	\overline{\varphi\vphantom{A}}_y &=& -\frac{\pi}{2}\,. \label{eq:phase-y-transient_average}
\end{eqnarray}
\section{Experiment \label{sec:experiment}}
\subsection{Experimental setup \label{sec:experimentalSetup}}
\begin{figure}
	\centering
  \includegraphics[width=0.6\columnwidth]{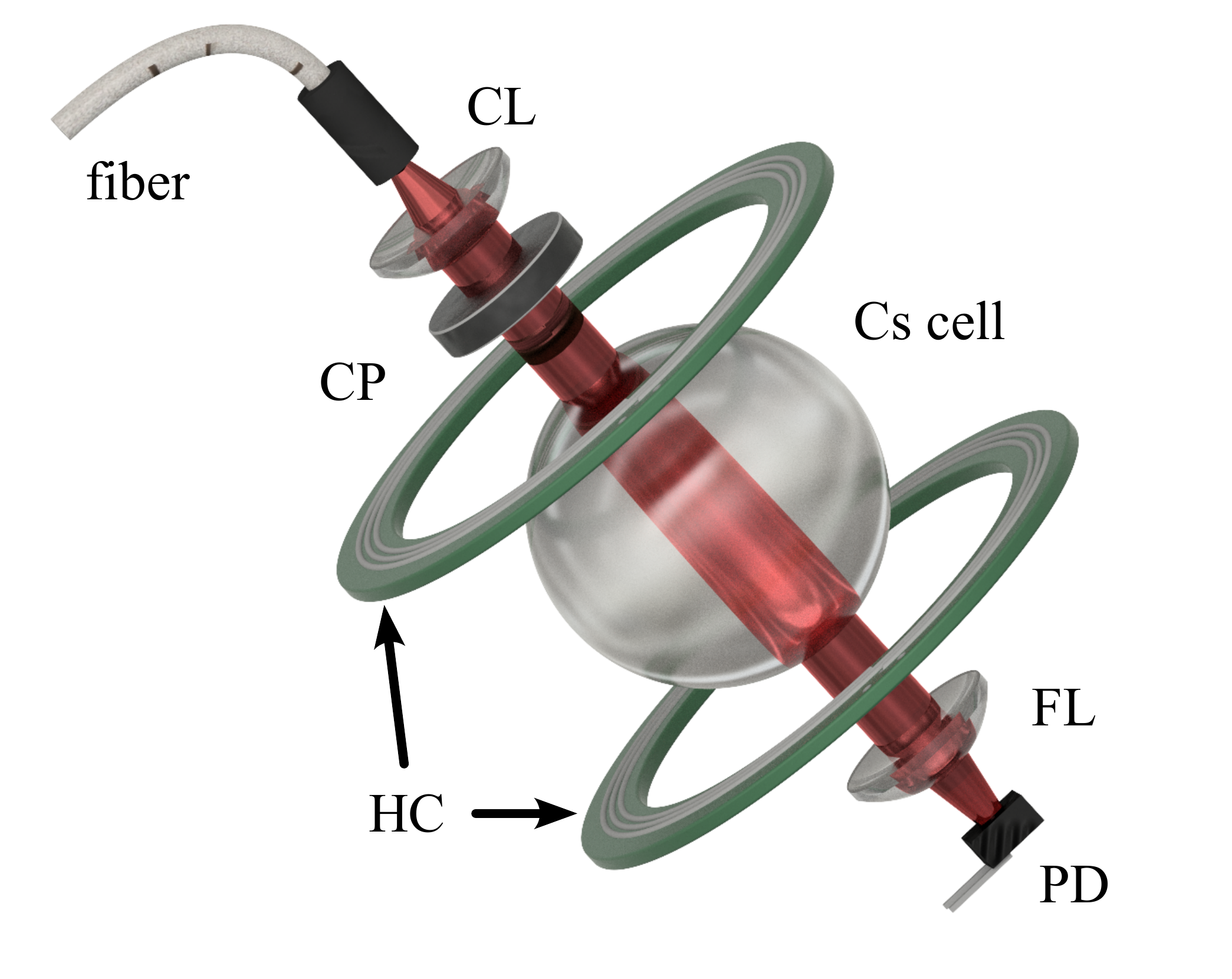}
	\caption{\label{fig:TSM-Real} Schematic of the TSM magnetometer head. The key components are illustrated; the plastic housing is not shown for clarity. CL – collimating lens, CP – circular polarizer, HC – Helmholtz coils, FL – focusing lens, PD – photodiode detector. See main text for details.
	}	
\end{figure}

For our TSM sensor head, we employ a $28\,\mathrm{mm}$ diameter spherical ${}^{133}\mathrm{Cs}$ vapor cell without buffer gas, featuring an inner anti-relaxation paraffin coating.
The cell operates at room temperature, includes a side arm serving as Cs reservoir, and features an intrinsic relaxation rate of $\approx 3~\mathrm{Hz}$ \cite{Castagna2009}.
A plastic mount, made by 3D-printing, encloses the cell and holds the fiber coupler, optical elements, photodiode, and the Helmholtz coils (HC).
The HC, constructed from a printed-circuit board (50~mm diameter, 25~mm separation, 6 windings each), apply the rf magnetic field $\vec{B}_\mathrm{rf}$ parallel to $\vec{k}$ (see Fig.~\ref{fig:TSM-Real}).
The amplitude and frequency of $\vec{B}_\mathrm{rf}$ are controlled by a lock-in amplifier (\textit{Zurich Instruments}, model \textit{HF2LI}).
We found that optimal TSM sensitivity is achieved with a 400~mV amplitude lock-in output connected to the HC via a 27~k$\ohm$ series resistor, resulting in $\Grf=2\pi\,3.95$~rad/s.
The same signal serves as the reference for lock-in demodulation.
Narrow-band ($\delta\nu_\mathrm{l} \leqslant 5\,\mathrm{MHz}$) laser light at $894.6\,\mathrm{nm}$ from a tunable diode laser system (\textit{Toptica Photonics}, model \textit{DL pro}) is delivered via a multi-mode fiber, collimated by a lens (CL), then circularly polarized by a circular polarizer (CP).
The circular polarizer consists of a polarizer coupled to a quarter-wave plate.
While traversing the cell, the laser light optically pumps the Cs vapor and its intensity is modulated by the precessing spin orientation $\vec{S}(t)$ driven by $\vec{B}_\mathrm{rf}$.
The cell-transmitted light is focused by a second lens (FL) and detected by a photodiode (PD).
The angle between the laser beam propagation direction $k$ and the static magnetic field $\vec{B}_0$ was set to $45\degree$, near the optimal sensitivity point while keeping the geometry simple.
The photo current generated in the PD is converted into voltage by a transimpedance amplifier (\textit{Femto Messtechnik}, model \textit{DLPCA-200}, gain $10^6$ V/A, -3dB bandwidth 200 kHz), and the resulting signal is demodulated by the lock-in amplifier.

\begin{figure}
	\centering
 \includegraphics[width=0.7\columnwidth]{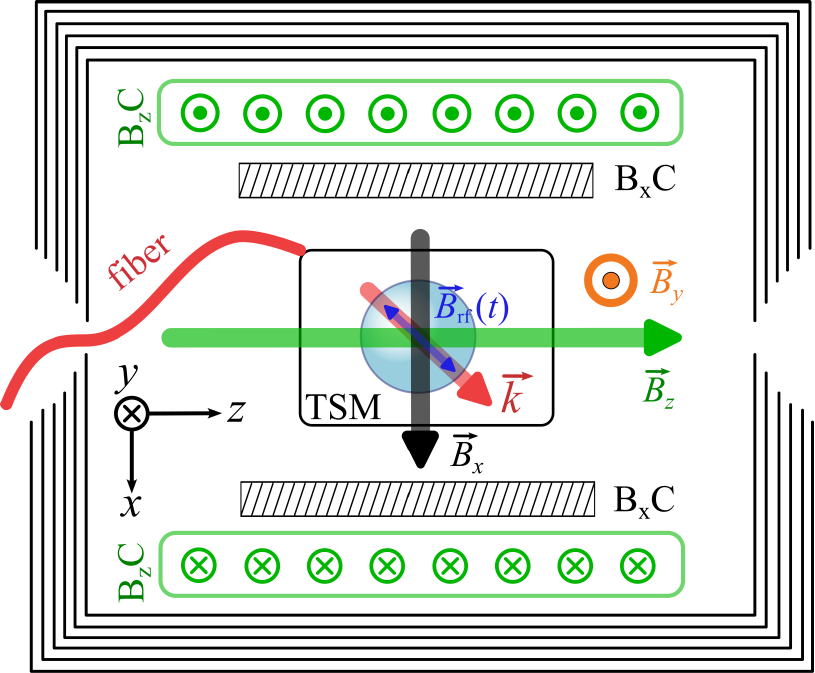}
	\caption{\label{fig:Setup}
        Experimental setup schematic. A detailed view of the TSM head is shown in Fig.~\ref{fig:TSM-Real}. The magnetic field is generated by a triaxial coil system in a Helmholtz configuration. The cylindrical coil for $B_z$ has 15 windings and a diameter of 220 mm, while the square coils for $B_x$ and $B_y$ have side dimensions of $d_x$ = 150 mm and $d_y$ = 170 mm, each with 15 windings.
	}	
\end{figure}

The TSM sensor head is placed within a cylindrical six-layer $\mu$-metal magnetic shielding (see Fig.~\ref{fig:Setup}).
A static field of $|\vec{B}_z|= 1.90\,\mathrm{\mu T}$ is applied along the $z$-axis using a solenoid driven by a low-noise current source (\textit{Koheron}, model \textit{DRV300-A-10}) at 9.19~mA.
Additional fields $\vec{B}_{x,y}$ are generated by pairs of Helmholtz coils located inside the solenoid.
We ensured the Cs vapor cell was positioned at the center volume of all coils, maximizing the magnetic field homogeneity.
To rotate $\vec{B}_0$ without changing its modulus $|\vec{B}_0|$, we apply a 50\% duty-cycle square wave to the respective additional coil.
For example, to tilt $\vec{B}_0$ within the $xz$-plane (around the $y$-axis), a bipolar square-wave current is applied to the $\vec{B}_{x}$ coil.
This creates a total field of $\vec{B}_0=(B_x,0,B_z)^{\top}$ during the positive semi-period and $(-B_x,0,B_z)^{\top}$ during the negative semi-period, maintaining a constant modulus of $\sqrt{B_x^2+B_z^2}$.
We developed a dedicated trigger circuit to synchronize the tilting moment with a specified phase of $\vec{B_\mathrm{rf}}$.  
This allowed direct measurement of the $\phi$ dependence as predicted by Eqs.~\ref{eq:phase_transient} and \ref{eq:no_phase_transient}.
The circuit consists of a Schmidt trigger and a delay circuit triggered by the rising edge of the sine wave coming from lock-in amplifier. 
The phase shift is controlled internally by the lock-in amplifier.

We observe the steady-state phase signal and tune the DC offset current through the $B_x$-coil to compensate for any remnant field components along the $x$-direction.
The same procedure is then applied to tip $\vec{B}_0$ within the $yz$-plane (around the $x$-axis) using the $B_y$-coil instead.
\subsection{Experimental results \label{sec:experimental_results}}
\begin{figure}
	\centering
	\includegraphics[width=0.85\columnwidth]{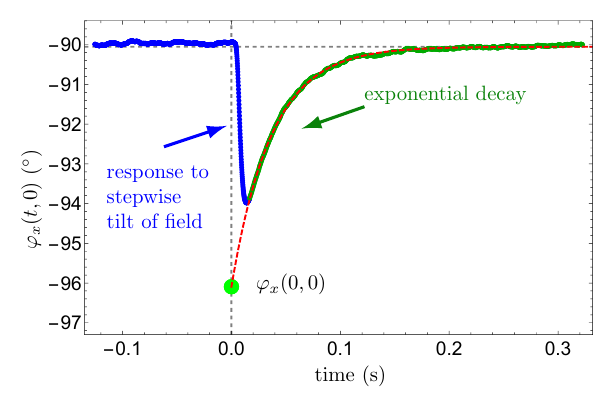}
	\caption{\label{fig:FitOfThePhase}
		Observed phase $\varphi_x(t,0)$ jump while rotating $\vec{B_0}$ for an angle of $\beta_x =2.36 \degree$ with $\phi = 0 \degree$.
		The blue line shows the phase before rotation and during transient process determined by low-pass filter characteristics of the lock-in amplifier.
        The blue data points represent the magnetometer's initial phase before rotation and response to a fast step-wise tilt of the magnetic field.
        The observed deviation from a step-wise response is due to the lock-in's low-pass filter time constant.
		The dark green data points present decaying phase after filter of the lock-in has settled.
		The red dashed line shows the fit of the data in dark green according to Eq.~\ref{eq:fit-varphi-t}.
		The fit function is extrapolated to $t=0$ (green point) to obtain $\varphi_{x}(0,0)$.
		The vertical dashed line at $t=0$ marks the moment when the rotation happens.
		The horizontal dashed line marks the asymptotic value of $\varphi \approx \pi/2$ after the transient process has ended.
		From the fit, we estimate $\gamma_\mathrm{tot}=2 \pi \, 4.3(3)$~rad/s and $a_x= -6.04(15) \degree$.
		}
\end{figure}

As an example, Figure~\ref{fig:FitOfThePhase} shows the phase $\varphi_x(t,0)$ recorded by the lock-in amplifier for $\beta_x=2.36 \degree$ and $\phi=0$.
We extract the amplitude $a_x$ by fitting the data to:
\begin{eqnarray}
	\varphi_{x,y}(t,\phi) &=&-\frac{\pi}{2} + o + a_{x,y}(\phi)\, e^{-\gamma_\mathrm{tot} t}\,,
	\label{eq:fit-varphi-t}
\end{eqnarray}
where $o$ accounts for an offset due to small Larmor frequency detuning, $a_{x,y}(\phi)$ is the initial amplitude of the signal, and $\gamma_\mathrm{tot}=2\pi \, 4.3(3)$~rad/s is the relaxation rate.
The total relaxation rate, $\gamma_\mathrm{tot}$, later used to compare theoretical prediction and the experimental observations, is estimated as a mean value of multiple fits for various values of $\phi$, with the uncertainty inferred from the resulting standard deviation.
The fit is performed on the dark green data points in Fig.~\ref{fig:FitOfThePhase}.
The blue data points represent the magnetometer's initial state before rotation and the lock-in's low-pass filter response to a transient (step) signal.

\begin{figure}
	\centering
	\includegraphics[width=0.9\columnwidth]{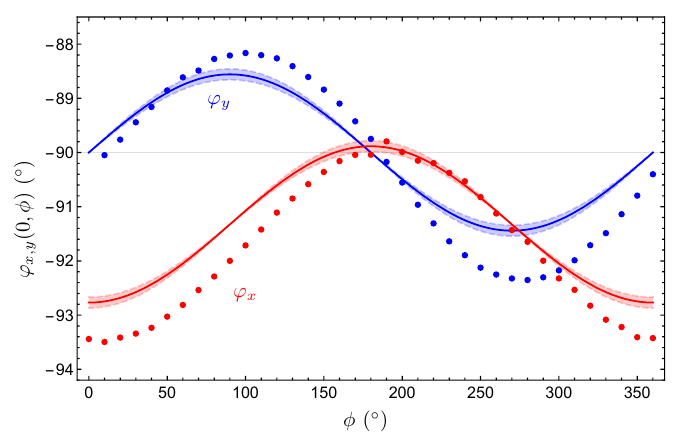}
	\caption{\label{fig:Amplitude-vs-phi} Comparison of theoretical prediction with experimental observations of the initial phase jump amplitude as a function of the rf field phase.
	Data points represent $\varphi_{x,y}(0,\phi)$ extracted from fits to Eq.~\ref{eq:fit-varphi-t}, while solid lines represent the theoretical predictions for rotation of $\vec{B_0}$ around $x$- (red) and $y$-axis (blue).
	The theoretical line was generated using $\beta_{x,y}=-1.33\degree$, $\gamma_\mathrm{tot}=2\pi\,4.3(3)$~rad/s and $G_\mathrm{rf}=2\pi\,3.96$~rad/s.
	The shaded area indicates the uncertainty range due the estimation error of $\gamma_\mathrm{tot}$.
	}
\end{figure}

We perform a series of such measurements for different $\phi$ and $\beta_{x,y}=-1.33 \degree$.
Figure~\ref{fig:Amplitude-vs-phi} shows the experimental results (points) and theoretical predictions (solid lines).
The theoretical fit function is:
\begin{eqnarray}
	\varphi_{x,y}(0,\phi) &=& -\frac{\pi}{2} + \theta_{x,y} + \beta_{x,y} C_{x,y} \sin(\phi+\phi_\mathrm{sh})\,.
	\label{eq:fit-varphi-phi}
\end{eqnarray}

In Fig.~\ref{fig:Amplitude-vs-phi}, red denotes rotation around the $x$-axis and blue denotes rotation around the $y$-axis.
The results confirm the predictions from Eqs.~\ref{eq:phase_transient} and \ref{eq:no_phase_transient}.
The small discrepancies may be attributed to the approximations made, potential misalignment of the magnetometer with respect to the field coils, and phase delays or shifts induced by the electrical circuitry. 

We performed three sets of measurements with $\beta \in \{-0.66\degree, -1.33\degree, -2.36\degree\}$ for various values of $\phi$, as described above and shown in Fig.~\ref{fig:Amplitude-vs-phi}.
We used the fit function Eq.~\ref{eq:fit-varphi-phi} to extract values for $\theta_{x,y}$ and $C_{x,y}=\gamma_\mathrm{tot}/G_\mathrm{rf}$.
Note that, according to Eq.~\ref{eq:amplitude-y}, $\theta_y$ should be zero, but our experimental data show otherwise.
The fit parameter $\phi_\mathrm{sh}$ accounts for a small phase shift and the transformation between sine and cosine functions.

\begin{figure}
	\centering
	\includegraphics[width=0.89\columnwidth]{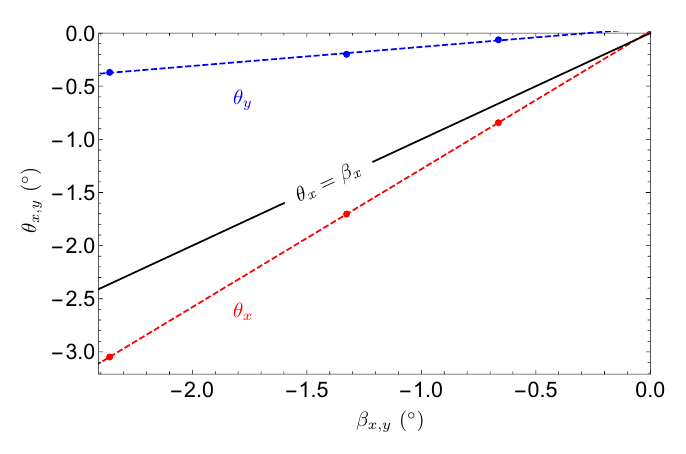}
	\caption{\label{fig:Theta-vs-beta} Extracted values of $\theta_{x,y}$ as a function of $\beta_{x,y}$.
    The data points represent the fit results (Eq.~\ref{eq:fit-varphi-phi}), dashed lines show linear regressions,
    and the solid black line represents the theoretical expectation $\theta_x=\beta_x$ as $\alpha=45\degree$ (see Eq.~\ref{eq:theta} for details).
	 }
\end{figure}

Figure~\ref{fig:Theta-vs-beta} shows the extracted values of $\theta_{x,y}$ as a function of $\beta_{x,y}$.
As $\alpha=\pi/2$ in our experiment, $\theta_x = \beta_x$ (presented as a solid black line), according to Eq.~\ref{eq:theta}.
The dots represent the results extracted from the fit, and the dashed lines represent linear regressions.
Both $\theta_{x,y}$ are close to the theoretical predictions. 
We attribute the small deviation to mechanical misalignment of the magnetometer with respect to the field coils, estimated to be on the order of a few degrees.

\begin{figure}
	\centering
	\includegraphics[width=0.85\columnwidth]{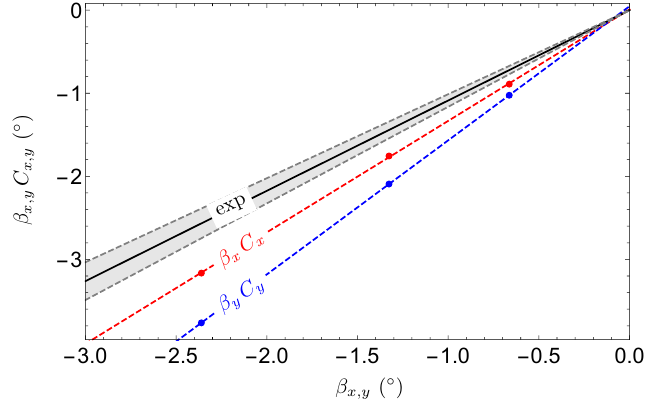}
	\caption{\label{fig:Amplitude-vs-beta}  Extracted values of $\beta_{x,y}\,C_{x,y}$ obtained from fits using Eq.~\ref{eq:fit-varphi-phi}.
		Data points represent the fit results, dashed lines show linear regressions, and the black solid line represents the expected (exp) result for $\alpha=45\degree$, $\gamma_\mathrm{tot}=2\pi\, 4.3(3)$~rad/s and $G_\mathrm{rf}=2\pi\,3.96$~rad/s. The shaded area indicates the uncertainty due to the fitted value of $\gamma_\mathrm{tot}$.}
\end{figure}

Figure~\ref{fig:Amplitude-vs-beta} shows the dependence of $\beta_{x,y}\,C_{x,y}$ on $\beta_{x,y}$.
The data points represent the results of a fit using Eq.~\ref{eq:fit-varphi-phi}, and the dashed lines show linear regressions.
The solid black line, with shaded area, represents the expected value of $\beta_{x,y}\,C_{x,y}$, including the uncertainty due to $\gamma_\mathrm{tot}=2\pi \, 4.3(3)$~rad/s.
\section{Discussion}
\label{sec:discussion}
The DPPE can be explained solely from classical Bloch spin dynamics and is unrelated to other known phenomena causing orientational dependence in nominally scalar OPM readout, such as 'classical' heading errors arising from nonlinear Zeeman splitting, (vector) light shifts, channel balancing in differential schemes, or dead zones, as recently discussed, e.g. in \cite{Oelsner2019}.
While 'classical' heading errors increase with increasing magnetic field modulus, the DPPE amplitude depends on tilting angle $\beta$, which, to first order is $\propto B_\perp /B_0$.
Thus, the effect cannot be mitigated by restricting operation to low magnetic fields.

Inspection of Eqs.~\ref{eq:phase_transient}-\ref{eq:amplitude-y} reveals a strong dependence of DPPE amplitude and transient duration on the total relaxation rate $\gamma_\mathrm{tot}$ and the strength of $B_\mathrm{rf}$ (i.e. $G_\mathrm{rf}$).
As shown in Eqs.~\ref{eq:SLxprime} and \ref{eq:SLyprime}, DPPE scales with $\gamma_p$, and thus, in a low-power regime, exhibits a linear dependence on light intensity.
This behaviour is similar to, but distinct from, (AC Stark) light shift phenomena.
However, unlike light shifts, DPPE does not require spectrally detuned light.

One strategy to avoid DPPE is to restrict applications to scenarios where variations of the magnetic field vector are parallel to $\vec{B}_0$.
Another strategy could be to limit magnetic field variations to the $x$-axis (rotation around the $y$-axis) and use averaging of many randomly timed samples, relying on Eq.~\ref{eq:phase-y-transient_average} to suppress DPPE by averaging over random $\phi$.

For rotation around the $x$-axis, we can estimate the required pointing stability of the magnetic field to keep the DPPE below typical noise and accuracy figures of an $M_x$ magnetometer:

For our ${}^{133}\mathrm{Cs}$ magnetometer with a total relaxation rate of $\gamma_\mathrm{tot} = 2\pi\, 4.3~\mathrm{rad/s}$ (as measured), the slope of the phase signal at its symmetry point is $\Delta\varphi/\Delta B= \gamma_{F}/\gamma_\mathrm{tot} = 0.81\,\mathrm{rad}/\mathrm{nT}\,$ (cp. Eq.~\ref{eq:phase}). 
This means a $1\,\mathrm{pT}$ field change in $|\vec{B}_0|$ corresponds to a change in phase signal of $\Delta\varphi = 46 \,\mathrm{mdeg}$.

With $G_\mathrm{rf}=2\pi\,3.96~\mathrm{rad/s}$, and $\alpha=45\degree$, according to Eq.~\ref{eq:phase_transient} to get an error due to DPPE of $\Delta\varphi=46\,\mathrm{mdeg}$ at $t=0$, the required tipping angle $\beta$ is only $21\,\mathrm{mdeg}$. 
In other words, DPPE means that $\vec{B}_0$ tipping angles of $\beta=0.0021\degree$ ($\beta=0.00021\degree$) around the $x$-axis will lead to transient readout errors resembling effective magnetic field changes of $100\,\mathrm{fT}$ ($10\,\mathrm{fT}$) in the phase signal of our magnetometer.
Achieving this level of magnetic field pointing stability represents a challenging technical task, even in very well-magnetically shielded environment.

This result gains further relevance when considered in the context of technical noise sources, such as magnetic field interference from power lines (50/60 Hz and harmonics).
The direction of these fields with respect to the static magnetic field is rarely known or well-controlled.
For example, an interference field component of $\vec{B}_\mathrm{IF}=18.3\,\mathrm{nT}$ perpendicular to the geomagnetic field vector ($|\vec{B}_0|=50\,\mathrm{\mu T}$ assumed here) would be sufficient to generate a 1\,pT amplitude DPPE.
This is on the order of values typically observed in a magnetically calm lab or field environment.

When the magnetometer is operated in free-running mode (see~Sec.~\ref{sec:introduction}), the phase signal will exhibit excursions that mimic changes in the static magnetic field modulus.
Beyond that, when a PLL is active with sufficiently large bandwidth to cover the DPPE dynamics, the feedback will lead to erroneous control of $\omega_\mathrm{rf}$.
This will introduce additional noise in the sensor readout and can represent an obstacle to increasing the sensor bandwidth by increasing the PLL bandwidth. 
To obtain measurements undisturbed by DPPE, the magnetometer needs to be allowed to settle into its steady state, assuming that the magnetic field tilts occur on a much slower time scale than the DPPE response.
However, in this case, as the response time is inversely coupled to the relaxation rate $\gamma_\mathrm{tot}$ (typically minimized to obtain high sensitivity), this time scale can be very long, severely limiting the sensor bandwidth.

This limitation to bandwidth represents a strong obstacle in applications where the sensor is moved and/or rotated in space, within the geomagnetic magnetic field, particularly when used on a moving platform.
Here, vapor cells featuring larger $\gamma_\mathrm{tot}$, e.g. (microfabricated) cells featuring buffer gas, are less affected, thus potentially advantageous.

DPPE is not restricted to single-beam $M_x$ magnetometers or schemes that detect a change in transmitted light intensity.
Moreover, we expect it to be relevant in any configuration where rf magnetic fields are used to coherently drive the spin precession of a spin-polarized ensemble, and a phase-sensitive read-out is employed.
This includes more sophisticated OPM pump-probe schemes, detection of nonlinear magneto-optical or Faraday rotation \cite{Budker2013}.
The effect is not restricted to thermal vapors; it could also be observed in other rf- or microwave-driven spin-polarized systems, such as ultracold atoms \cite{Cohen2019}, liquids or solid-state systems.

In mobile or rotating platforms, where $\vec{B}_0$ undergoes continuous reorientation with respect to the sensor, the magnitude and timescale of the DPPE effectively set a lower bound on the short-term stability of the magnetometer.
Beyond these immediate implications for PLL feedback operation and field-tracking accuracy, the DPPE phenomenon also has broader relevance in the context of quantum navigation and inertial sensing.  
In such applications, OPMs operated within the geomagnetic field are used to reconstruct position, heading and/or attitude information.
Therefore, any transient deviation of the lock-in phase translates into a fictitious heading-dependent bias field, which in turn limits the achievable measurement resolution.  

The effect is equally relevant in precision metrology as well as in fundamental physics experiments, where dc or low-frequency phase errors are often indistinguishable from genuine frequency shifts of physical origin.  
In those scenarios, DPPE constitutes in an intrinsic systematic contribution, originating from the vectorial light–atom interaction that underlies the optical detection process.  
The phenomenon is therefore not merely a technical imperfection of particular sensor geometries, but it also reflects a fundamental limitation that is associated with the finite relaxation rate and non-collinear projection of the driven spin components during the reorientation of $\vec{B}_0$. 
\section{Conclusions \label{sec:conclusion}}
We investigated the demodulation phase signal response in rf-driven optically pumped magnetometers.
The dependence of the static phase offset on the magnetic field vector’s direction relative to the sensor orientation, as observed in classical $M_x$ magnetometers, can be largely overcome by aligning the rf field oscillation and light propagation direction (TSM). 
However, while the static phase projection error (SPPE) is largely absent in the TSM configuration, we demonstrate that even in this scheme a transient phase signal arises from tilts of the static magnetic field relative to the sensor orientation (DPPE). 
We obtained analytical results by solving the modified Bloch equation which are in agreement with our measurements using an $M_x$ magnetometer employing a ${}^{133}\mathrm{Cs}$ paraffin-coated vapor cell.
We discuss the importance of the effect in practical applications due to its consequences regarding the static magnetic field pointing stability and implications on sensor bandwidth.
The results highlight fundamental limitations of rf-driven implementations of optically pumped magnetometers, which, despite their wide-spread use in research, have not been discussed in literature to date. 
Alternative approaches, based on the observation of free spin precession, as presented, e.g., in  \cite{Grujic2015,Afach2015,Hunter2018,Gerginov2020}, are not susceptible to DPPE and therefore offer advantages in terms of accuracy and rotation invariance. 
In our opinion, this, alongside technical advances in required sensor control and readout electronics, provides a strong rationale for the renewed interest in these alternative pulsed OPM designs.
\backmatter
\section*{Declarations}
\subsection*{Funding}
M.~Ć., A.~K., Z.~G., and T.~S. acknowledge support from the German Federal Ministry of Research, Technology and Space (BMFTR) within the project “free alignment precession optically pumped magnetometer (FRAPOPM)” (grant number 01DS21006).
M.~Ć., A.~K., Z.~G. also acknowledge funding provided by the Institute of Physics Belgrade through a grant from the Ministry of Education, Science and Technological Development of the Republic of Serbia.
T.~S. further acknowledges support within the project Rio-GNOME funded by the German research foundation (DFG) under grant number 439720477.
\subsection*{Competing interests}
The authors declare that they have no competing interests.
\subsection*{Availability of data and materials}
The datasets and the code used and/or analysed during the current study are available from the corresponding author on reasonable request.
\subsection*{Authors' contributions}
Z.~G.: Conceptualization, Methodology, Formal analysis, Investigation, Visualization, Validation, Writing - Original Draft, Writing - Review \& Editing, Supervision.
M.~Ć.: Investigation, Validation, Writing - Review \& Editing.
A.~K.: Investigation, Writing - Review \& Editing.
A.~W.: Conceptualization, Methodology, Resources, Writing - Review \& Editing.
T.~S.: Conceptualization, Methodology, Formal analysis, Validation, Writing - Original Draft, Writing - Review \& Editing, Funding acquisition, Resources, Project administration, Supervision.
\subsection*{Acknowledgements}
Not applicable.
\end{document}